\documentclass[pra,twocolumn]{revtex4}

\usepackage{amsmath}
\usepackage{amssymb}
\usepackage{graphicx}
\usepackage{times}

\renewcommand{\H} {{\cal H}}
\newcommand{\summ} {\sum_m}
\newcommand{\X} {\Upsilon}
\newcommand{\gnu} {\gamma+\nu}
\newcommand{\keff} {\kappa_{\mathrm{eff}}}
\newcommand{\Deff} {D_{\mathrm{eff}}}
\newcommand{\Pb} {{\overline{\Pi}}}
\newcommand{\Sb} {{\overline{\Sigma}}}
\newcommand{\Lb} {{\overline{\Lambda}}}

\begin{document}
\author{Thomas Salzburger and Helmut Ritsch}
\title{Lasing and cooling in a hot cavity}
\affiliation{Institute for Theoretical Physics, University Innsbruck\\
  Technikerstrasse 25/2, 6020 Innsbruck, Austria}

\date{\today}

\begin{abstract}
We present a microscopic laser model for many atoms coupled to a single cavity mode, including the light forces resulting from atom-field momentum exchange.
Within a semiclassical description, we solve the equations for atomic motion and internal dynamics to obtain analytic expressions for the optical potential and friction force seen by each atom.
When optical gain is maximum at frequencies where the light field extracts kinetic energy from the atomic motion, the dynamics combines optical lasing and motional cooling.
From the corresponding momentum diffusion coefficient we predict sub-Doppler temperatures in the stationary state.
This generalizes the theory of cavity enhanced laser cooling to active cavity systems.
We identify the gain induced reduction of the effective resonator linewidth as key origin for the faster cooling and lower temperatures, which implys that a bad cavity with a gain medium can replace a high-Q cavity.
In addition, this shows the importance of light forces for gas lasers in the low-temperature limit, where atoms can arrange in a periodic pattern maximizing gain and counteracting spatial hole burning.
Ultimately, in the low temperature limit, such a setup should allow to combine optical lasing and atom lasing in single device.    
\end{abstract}

\maketitle

\section{Introduction}
Light forces in optical resonators are strongly modified as compared to free space \cite{Pinkse,Hood,Hechenblaikner}.
Extending the ideas of laser cooling in free space, various new schemes of laser cooling of atoms have now been developed in the field of cavity QED theoretically (for a review, see e.g.\ Ref.~\cite{cavcoolrev}) and experimentally \cite{Kuhn}. 
The common basis of all these considerations constitute the strong correlations between the atomic motion and the dynamics of the cavity field.
Contrary to conventional laser cooling schemes relying on spontaneous emission, dissipation of energy and entropy from the atomic motion is provided here by the cavity decay channel, where the emission of blue shifted photons is enhanced.
Under ideal circumstances, it is possible to implement very efficient laser cooling schemes with strongly suppressed spontaneous emission.
Hence, the underlying mechanism is applicable to any polarizable particle and, in particular, also to molecules where no closed excitation cycle is necessarily available.
As momentum diffusion (heating) originates mainly from field fluctuations due to the randomness of photon loss via the cavity mirrors, the attainable atomic temperature is limited by the cavity linewidth $\kappa$ \cite{Horak,Hechenblaikner}.
In high-finesse cavities, one even gets sub-Doppler cooling \cite{Rempe}, but at the expense of a slow cooling rate.

Among several suggestions to circumvent this slowdown, like e.g.\ using collectively enhanced light scattering \cite{Black}, it has been recently proposed to add a gain medium into the cavity to achieve lower temperatures and fast cooling simultaneously \cite{Vuletic}.
The central idea here amounts to obtain an effectively small cavity linewidth via intracavity amplification in a not so good resonator.
In a first approach to this idea, Vuletic predicted improved cooling by a strong reduction of the effective cavity linewidth through addition of intracavity gain \cite{Vuletic}.
In essence, the partial compensation of the cavity round trip loss by gain gives rise to an enhancement of the cavity-to-free-space scattering ratio substantially above unity, even for rather lossy resonators.
As, in such systems, the cooling force is proportional to this scattering ratio, he predicted both fast cooling and low final atomic temperatures.
Of course, gain in quantum mechanics is inevitably connected with fluctuations, so that the validity of the connection of temperature and linewidth is not so obvious any more in this case.

In the present work, we follow this general idea but use a more refined approach based on microscopic modeling of gain by an ensemble of inverted atoms.
The effects of quantum fluctuations and saturation are thus automatically modeled in a self-consistent way and the model is a direct generalization of our recent investigations of a single-atom--single-mode laser including light forces \cite{SALletter,SALarticle}.
In this configuration, we could show that by a suitable choice of parameters (i.e.\ blue detuning of the field mode with respect to the atomic transition), single-atom lasing can be combined with trapping and cooling of the atom at a cavity field mode antinode.
In the stationary limit, the average kinetic energy of the particle can be lower than the optical potential depth of the laser field that is created by the atom itself.
Extending this laser model to many atoms, we directly get a microscopic model for a hot cavity by simply fixing the positions of some of the atoms.
These represent our gain medium as they are partially inverted through external driving.
Of course, beyond merely providing for gain, they play a much more complex role in the combined atom-field dynamics.

From a different point of view, we are extending the standard Haken-Lamb laser model \cite{Lamb,Haken} by including light forces on the active atoms and study lasing as well as their trapping and cooling properties.
As compared to the single-atom laser model, the trapping field is now generated from all the active atoms simultaneously, so that each one has to be less inverted to achieve a desired optical potential (photon number).
This should reduce spontaneous emission and momentum diffusion of the atoms and we can expect better trapping and lower temperatures.
In a first step to simply mimic an active medium, we will keep all atoms but one fixed at antinodes and study the dynamics of the remaining atom.
In the parameter regimes where we find trapping and cooling of the test atom, its parameters can then be prescribed to the other atoms to find a self-consistent final solution for all atoms. 

Our work here is organized as follows:
in Sec.~\ref{sec:model} we present the general model and define the relevant quantities and parameters.
In Sec.~\ref{sec:semi} we numerically study the semiclassical limit of the equations where we replace the field and atomic operators by their expectation values and factorize expectation values of operator products.
A laser model which includes second order quantum correlations is presented and analytically solved in Sec.~\ref{subsec:active}.
The results are then used as a hot-cavity description for the motion of an extra atom in Sects.~\ref{subsec:motion} - \ref{subsec:scaling}.
Again we can find analytic expressions for the relevant physical quantities allowing a thorough discussion of the properties of the present system.
In particular, we focus on the dipole force and the optical potential for a fixed atom in Sec.~\ref{subsec:steady}.
In Sec.~\ref{subsec:temperature}, we calculate the friction and diffusion coefficients which give rise to an estimate of the atomic localization and equilibrium temperature.
Finally, the scaling properties of these quantities with the number of atoms are discussed in Sec.~\ref{subsec:scaling}.

\section{model}
\label{sec:model}
In this section, we briefly review key elements of the Haken-Lamb laser model \cite{Haken,Lamb}, which ignores atomic motion, to get a first insight into the dynamics on a simple and intuitive level.
The laser active gas consists of $M$ identical two-level atoms (represented by their Pauli operators $\sigma_{-m}$, $\sigma_{+m}$, and $\sigma_{z,m}$) which interact with a single mode supported by an optical high-finesse resonator.
Each atomic dipole is coupled to the cavity mode (represented by the bosonic field operator $a$) with coupling constant $g$ times the mode function $f(x)$ evaluated at the corresponding position of the atom $x_m$.
In the simplest (one-dimensional) case, the mode function is given by $f(x) = \cos(k x)$ with mode wavenumber $k$.
The cavity resonance frequency $\omega_\mathrm{c} = kc$ is generally detuned from the atomic transition frequency $\omega_\mathrm{a}$ by a small amount $\Delta = \omega_\mathrm{c}-\omega_\mathrm{a}$, on the order of a few atomic linewidths at most.
In a frame rotating at the cavity frequency, the atom-field Hamilton operator including interaction then reads within the rotating-wave approximation
\begin{align}
  \label{H}
  \H_\mathrm{int} = &-\Delta\summ\sigma_{+m}\sigma_{-m} \nonumber\\
  &- i g\summ f(x_m)(a^\dagger\sigma_{-m} - \sigma_{+m} a)\,.
\end{align}
In addition to this coherent dynamics, the atoms as well as the resonator mode are also weakly coupled to the environment.
Using a standard master-equation approach \cite{Gardiner} for spontaneous emission of the atoms into the vacuum (with rate $2\gamma$ for each atom) and photon loss via the mirrors (with rate $2\kappa$), we get
\begin{equation}
  \label{me1}
  \dot{\varrho} = -i\left[\H_\mathrm{int},\varrho\right] + \mathcal{L_\gamma}\varrho + \mathcal{L_\kappa}\varrho\,,
\end{equation}
where the damping operators are given by
\begin{subequations}
  \begin{gather}
    \label{me1b}
    \mathcal{L}_\gamma\varrho = \gamma\summ\left(2\sigma_{-m}\varrho\sigma_{+m}
      - \left\{\sigma_{+m}\sigma_{-m},\varrho\right\}\right)\,,\\
    \mathcal{L}_\kappa\varrho = \kappa\left(2 a \varrho a^\dagger - \left\{a^\dagger a,\varrho\right\}\right)\,.
  \end{gather}
\end{subequations}

The steady state of this system so far is obviously a vacuum cavity field with all atoms in the ground state.
Here, we are interested in an active system and, since gain requires atomic population inversion, we need to add pumping of the atoms.
In practice, such pumping is achieved via an auxiliary atomic level actively coupled to the lower level which decays predominantly into the atomic upper state within a short time.
As introduced by Haken already some decades ago \cite{Haken}, this process can be modeled in a quantum mechanically consistent way by a so-called inverted heat bath approach which properly accounts for the fluctuations that are accompanied by gain, without the need to enlarge the atomic Hilbert space.
Basically, it amounts to introducing a spontaneous absorption rate $2\nu$, analogous to spontaneous emission but in the opposite direction.
Mathematically, we simply have to add the following Liouvillian term
\begin{equation}
  \label{me2}
  \mathcal{L}_\nu\varrho = \nu\summ\left(2\sigma_{+m}\varrho\sigma_{-m} - \left\{\sigma_{-m}\sigma_{+m},\varrho\right\}\right)
\end{equation}
to our master equation (ME).
Putting all terms together, the resulting ME is equivalent to the Heisenberg-Langevin equations (HLE) [with $G_m = g f(x_m)$],
\begin{subequations}
\label{hle}
  \begin{gather}
    \dot{a} = - \kappa a + \summ G_m\sigma_{-m} + \xi_a \,,\\
    \dot{\sigma}_{-m} = (i\Delta - \gamma - \nu)\sigma_{-m} + G_m\sigma_{z,m}a + \xi_{\sigma,m} \,,\\
    \dot{\sigma}_{z,m} = -2(\gamma+\nu)\sigma_{z,m} - 2G_m\left(a^\dagger\sigma_{-m} + \sigma_{+m}a\right) \nonumber \\
      + 2(\nu-\gamma) + \xi_{z,m}\,.
  \end{gather}
\end{subequations}
Fluctuations are properly included here in the noise terms $\xi_i$ which follow from Eqs.~\eqref{me1} and \eqref{me2}.
Notice, that at this point the atomic center-of-mass (COM) positions are simply contained as c-number parameters \cite{Cohen}.
As these equations are coupled and nonlinear, a direct analytical solution nevertheless seems out of reach and a vast amount of theoretical work on laser theory has been devoted to their solution in the past decades \cite{Lamb}.
Here, we will only resort to the most common proven approximations in the following.

\section{Semiclassical laser equations with light forces}
\label{sec:semi}
Starting from the corresponding quantum HLE for the polarization and field dynamics, we obtain approximate c-number equations by simply taking expectation values of these equations and factorizing expectation values of operator products into products of single-operator expectation values.
Within this approximation, we do not need to further specify the noise terms as their expectation values are zero by construction.
This approach is generally known as semiclassical laser theory.
The replacement $\left(\langle a\rangle,\langle\sigma_{-m}\rangle,\langle\sigma_{z,m}\rangle\right) \rightarrow \left(\alpha,s_m,z_m\right)$ then yields
\begin{subequations}
  \label{eqLamb}
  \begin{gather}
    \dot{\alpha} = -\kappa\alpha + \summ G_m s_m\,,\\
    \dot{s}_m = (i\Delta-\gamma-\nu)s_m + G_m z_m\alpha\,,\\
    \dot{z}_m = -2(\gamma+\nu)z_m - 2G_m(\alpha^*s_m + s^*_m\alpha) + 2(\nu-\gamma)\,.
  \end{gather}
\end{subequations}
In standard approaches, Eqs.~\eqref{eqLamb} contain the positions of the atoms as fixed parameters via $G_m=g f(x_m)$.
As a new element, we now introduce the atomic COM motion into the system so that the positions $x_m(t)$ get dynamical and are governed by light forces.
Within the point particle model which is based upon a classical treatment of the atomic COM motion \cite{semiclassical,Cohen}, the corresponding Newtonian equations of motion are
\begin{subequations}
  \label{external}
  \begin{gather}
    \dot{x}_m = p_m/m_\mathrm{at}\,,\\ 
    \dot{p}_m = F_m\,,
  \end{gather}
\end{subequations}
with the force acting on the $m$th atom given by
\begin{equation}
  F_m = -2 \, \mathbf{Im}\{\alpha^*s_m\} \left(\nabla_m G_m\right)
\end{equation}
and $m_\mathrm{at}$ denoting the atomic mass.
Note that this description of the external dynamics via point particles is only valid as long as the respective atomic kinetic energies are well above the recoil limit $E_\mathrm{rec} = \hbar^2k^2/(2m_\mathrm{at})$, which is well fulfilled down to atomic temperatures in the order of only a few micro Kelvin.

Although the particles are now moving, the system can still posses a quasistationary state.
Here the field as well as the atomic momentum and position distributions which might be characterized by an effective kinetic temperature are approximately constant in time.
As in other single-mode--two-level laser models, we find a steady-state energy balance relation for the total atomic ground and excited state populations [i.e.\ $P_\mathrm{g} = \summ(1-z_m)/2$ and $P_\mathrm{e} = \summ(1+z_m)/2$, respectively] and the mean photon number $N = |\alpha|^2$,
\begin{equation}
  \label{relLamb}
  \nu P_\mathrm{g} = \gamma P_\mathrm{e} + \kappa N\,.
\end{equation}
(Note that $P_\mathrm{g} + P_\mathrm{e} = M$.)
This relation describes the balance of energy fed into the system by the pumping mechanism and the photon loss via atomic spontaneous emission and cavity field decay.
Relation \eqref{relLamb} is quite universal and valid beyond the semiclassical treatment so that we will recover it in the quantum model in Sects.~\ref{subsec:active} and \ref{subsec:steady}.
Note that it does not include the kinetic and potential energy fed into the atomic motion in single emission or absorption events (recoil energy) since it is only a negligible fraction of the photon energy or internal atomic excitation energy and, thus, only very weakly modifies the conditions for the lasing threshold.
Thus the system should exhibit coherent oscillation if
\begin{equation}
  \label{threshold}
  w \summ f^2(x_m) = \kappa\frac{\gamma+\nu}{\nu-\gamma}\,,
\end{equation}
where we have introduced the rate of emission into the resonator mode per atom as
\begin{equation}
  w = \frac{(\gamma+\nu)g^2}{(\gamma+\nu)^2+\Delta^2}\,.
\end{equation}
Of course, the total rate $w_\mathrm{tot} = w \summ f^2(x_m)$ depends on the atomic positions and hence the spacial distribution of the atoms has influence on whether threshold is reached or not.
In the limiting case where all atoms are located close to field nodes, the total rate $w_\mathrm{tot}$ gets very small, which will eventually prevent the system from reaching threshold.
In this case, the single-atom excitation is determined by the rates of pumping and spontaneous emission, $\nu/(\gamma+\nu)$, and the cavity mode remains unpopulated.
Hence we also get not forces on the atoms, and the optical potential vanishes. 
On the other hand, when the system operates above threshold, the steady-state of the population inversion for each atom is fixed by
\begin{equation}
  w\summ f^2(x_m) z_m = \kappa\,,
\end{equation}
and the mean photon number can be determined from the relation \eqref{relLamb}.

\begin{figure}
  \includegraphics[width=8cm]{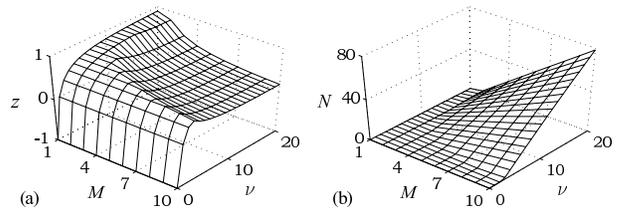}
  \caption{Surface plots of the common population inversion $z$ (left) and the photon number $N = |\alpha^2|$ (right) as a function of the atom number $M$ and the pumping rate $\nu$ in the stationary state for the situation where all atoms are fixed at field antinodes.
    The parameters are $(\gamma,\Delta,g) = (1,20,5)\kappa$.}
  \label{fig:eins}
\end{figure}
This behavior is depicted in Fig.~\ref{fig:eins} (a) where we plot the population inversion $z$ as a function of the atom number $M$ and the pumping strength $\nu$ in the case of lowest possible threshold with all atoms fixed at mode antinodes ($z_m = z$).
With growing $\nu$, the population inversion reaches its maximum value at threshold and eventually drops back to zero, when the growing stimulated emission (gain) decreases the upper state population.
The cavity photon number, on the other hand, increases linearly with both $\nu$ and $M$ as shown in Fig.~\ref{fig:eins} (b).

Let us now investigate the effect of light forces on the dynamics.
In this context it is interesting to note that for inverted atoms, the effect of intensity gradients is reversed as compared to thermal atoms.
Hence, as discussed in Refs.~\cite{SALletter,SALarticle}, an inverted atom is a high-field seeker for a blue detuned field.
As a consequence, we have to work with a blue detuned cavity, i.e.\ $\Delta > 0$, if we want our gain atoms to be drawn towards field antinodes where the gain is large.
Luckily, this coincides with the parameter regime where the motion can be cooled via gain as some kinetic energy is needed to optimally fulfill the energy resonance condition for stimulated emission \cite{SALletter}.

\begin{figure}
  \includegraphics[width=8cm]{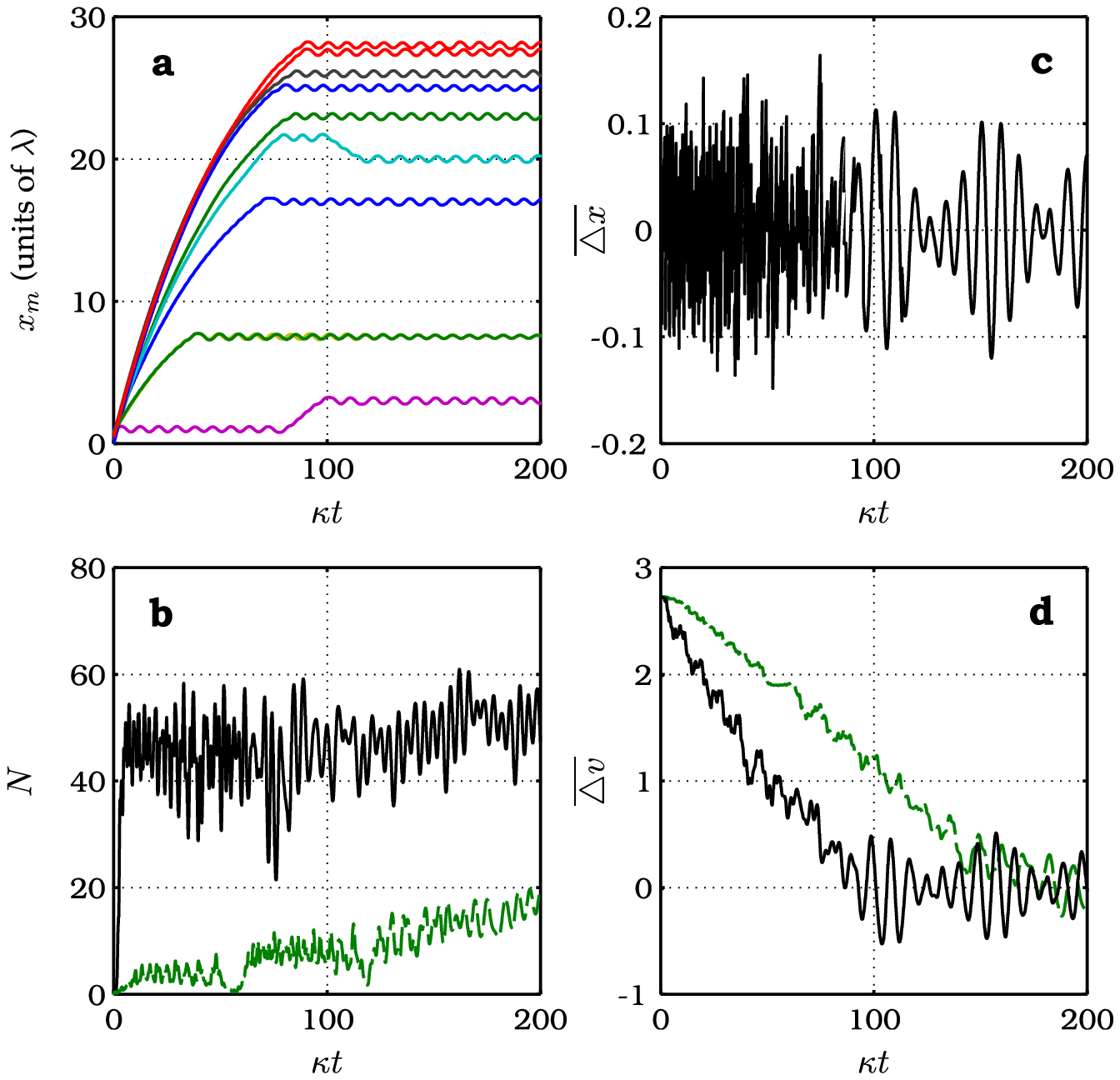}
  \caption{Sample trajectories obtained by numerically integrating Eqs.~\eqref{eqLamb} and \eqref{external} for 10 atoms.
    The parameters are the same as in Fig.~\ref{fig:eins} with $\nu = 20\kappa$.\\
    (a) Atomic positions in units of the wavelength $\lambda$.\\
    (b) Cavity photon number $N$.\\
    (c) Mean atomic distance to the nearest antinodes in units of $\lambda$.\\
    (d) Average of the atomic velocities in units of the Doppler velocity.\\
    The dotted lines in (b) and (d) correspond to the results for four atoms.}
  \label{fig:zwei}
\end{figure}

To demonstrate this we plot in Fig.~\ref{fig:zwei} (a) the trajectories obtained from a numerical simulation of Eqs.~\eqref{eqLamb} and \eqref{external} for an ensemble of ten atoms where the atoms start at random positions with random initial velocities.
The atoms are decelerated and eventually get trapped in the vicinity of field antinodes where they oscillate in single wells.
Inspection of Fig.~\ref{fig:zwei} (d) which shows the corresponding mean velocity truly reveals that the atomic motion is strongly damped at the beginning.
After the atoms get trapped they are still cooled further, although on a much slower timescale.
For comparison, we have added in Fig.~\ref{fig:zwei} (d) the curve for lowest possible atom number to pass threshold (dotted line) which is four for the chosen parameters.

Hence we see that, in principle, gain and cooling can simultaneously occur and that cooling can be significantly enhanced by the presence of other atoms in the same mode.
For favorable parameters, all the atoms have accumulated at the antinodes of the mode after some time and the system shows steady lasing at the maximum intensity possible.
The classical model, however, leaves out the fluctuations of the light forces and hence cannot be used to predict the final temperature of the atomic gas and the long term stability of this lasing process.
A consistent description then requires to include higher order quantum correlations of the operators and forces as presented in the next section.  

\section{Quantum description of cooling in a hot cavity}
\label{sec:quantum}
\subsection{Active cavity model}
\label{subsec:active}
In this section, we include higher order quantum corrections to the laser model presented in the previous section.
Let us start with all atoms located at the field antinodes as found in the long time limit above and define the collective second-order operators
\begin{subequations}
  \label{op_coll}
  \begin{align}
    \overline{\Pi} = & \frac{1}{M}\summ\sigma_{+m}\sigma_{-m}\,,\\
    \Sb = & \frac{1}{M}\summ \left(a^\dagger\sigma_{-m}+\sigma_{+m}a\right)\,,\\
    \Lb = & \frac{-i}{M}\summ \left(a^\dagger\sigma_{-m}-\sigma_{+m}a\right)\,,
  \end{align}
\end{subequations}
such that we can rewrite the interaction Hamiltonian \eqref{H} in the form
\begin{equation}
  \H_\mathrm{int} = -M\Delta\overline{\Pi} - Mg\overline{\Lambda}\,.
\end{equation}

Again, we can derive HLE for these operators compatible to the ME \eqref{me1} and \eqref{me2},
\begin{subequations}
  \label{HLEactive}
  \begin{align}
    \dot{\Phi} = & -2\kappa\Phi + Mg{\Sb} +\X_\Phi \label{phi} \,,\\
    \dot{\Pb} = & - 2\gamma{\Pb} - g{\Sb} + 2\nu\left(1-{\Pb}\right)  + \X_{\Pb} \label{pb} \,,\\
    \dot{\Sb} = & -\Gamma{\Sb} - \Delta{\Lb} + i M g\left[\Sb,\Lb\right] + \X_{\Sb} \label{activeSigma} \,,\\
    \dot{\Lb} = & -\Gamma{\Lb} + \Delta{\Sb} + \X_{\Lb} \label{lb}\,.
  \end{align}
\end{subequations}

It turns out that the collective operators defined in Eqs.~\eqref{op_coll}, together with the photon number operator $\Phi = a^\dagger a$, almost form a linear, closed set of equations.
The only terms leading out of the set are atomic cross correlations of the form
\begin{equation}
    \langle\sigma_{+m}\sigma_{-n}\rangle_{m\neq n}\,
\end{equation}
as well as the commutator in Eq.~\eqref{activeSigma}.
It is generally well established that the atom-atom correlations turn out to be small at least for sufficient distant atoms ($d\gg\lambda$) which see uncorrelated vacuum fluctuations \cite{Haake}.
In Eqs.~\eqref{HLEactive} we have introduced second-order noise operators that originate from the coupling of the system to the environment.
Later we will need expectation values of their autocorrelation functions which are directly related to the damping (diffusion) terms in the ME.  

Looking at the equations for the expectation values of the photon number operator $N=\langle\Phi\rangle$ and the mean upper state population $\overline{P} = \langle\Pb\rangle$, we recover the steady-state energy balance equation introduced in the previous section by setting the time derivatives to zero and adding Eqs.~\eqref{phi} and \eqref{pb}.
This relation now reads
\begin{equation}
  \label{relColl}
  M\nu(1-\overline{P}) = \kappa N + M\gamma\overline{P}
\end{equation}
and allows to determine the steady-state photon number.
Equations \eqref{HLEactive} are exact but still nonlinear as they incorporate an operator product in Eq.~\eqref{activeSigma}.
Due to this nonlinearity, the corresponding equations for the expectation values are not closed and have no explicit analytic solution.
Analogous to the semiclassical case, in principle, an additional equation for $\langle[\Sb,\Lb]\rangle$ is needed which in turn will involve higher order moments resulting in an infinite hierarchy of equations.
At this point we will again use a factorization approximation for higher order products as introduced already in earlier laser models \cite{Protsenko,cavcoolrev}.
It relies on the replacement $(2\Pb-1)\Phi\rightarrow\langle 2\Pb-1\rangle\Phi$ such that
\begin{equation}
  \label{apprColl}
  i M\left[\Sb,\Lb\right] \approx  2 z_M \Phi + 2 \Pb\,.
\end{equation}
Here, $z_M$ is a scalar parameter denoting the average single-atom population inversion.
It has to be determined self-consistently during the calculations of the steady state.
This assumption will drop part of the population-intensity correlations by applying approximation \eqref{apprColl} while it keeps the quantum correlations between the atomic polarization and field which are pronounced in lasers.

\begin{figure}
  \includegraphics[width=8cm]{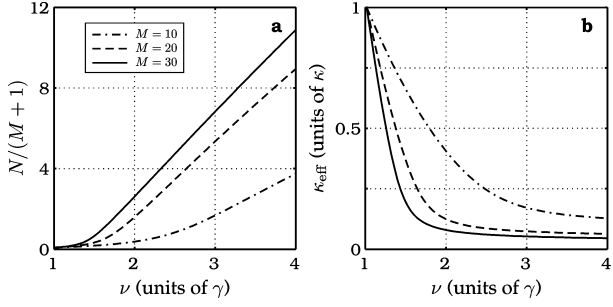}
  \caption{(a) Steady-state photon number per atom $n$ vs pumping rate $\nu$ for different atom numbers $M$, as indicated.
    The parameters are $(\gamma,\Delta,g) = (10,40,4)\kappa$.\\
    (b) Corresponding effectively reduced linewidth $\keff$.}
  \label{fig:drei}
\end{figure}

In a next step, we make use of the small linewidth of the cavity with respect to the atomic decay rates, which allows us to adiabatically eliminate the operators $\Pb$, $\Sb$ and, $\Lb$ such that we remain with a single differential equation for the photon number operator,
\begin{equation}
  \label{singleDiff}
  \dot{\Phi} =  -2\kappa(1-\zeta)\Phi + 2M\eta + \X_\Phi + \X_{\overline{\Phi}}\,.
\end{equation}
From this equation we see that the atoms introduce a gain term for the cavity field with an effective strength given by
\begin{equation}
  \eta = \frac{\nu w}{\gnu+w}\,.
\end{equation}
In this way, they compensate the round-trip loss of the resonator light and induce laser oscillation.
The gain parameter $\zeta$, defined by
\begin{equation}
  \label{zeta}
  \zeta = M\frac{(\gnu)w}{\kappa(\gnu+w)}z_M\,,
\end{equation}
is proportional to the population difference $z_M$ and is positive for positive $z_M$ (population inversion).
In this case, the medium effectively reduces the effective resonator damping according to
\begin{equation}
  \label{kappaEff}
  \keff = \kappa(1-\zeta)\,.
\end{equation}
Since the cavity-induced cooling force in a passive cavity is inversely proportional to the loss rate, we already can expect the cooling efficiency to be enhanced for $\zeta>0$.
On the other hand, since the atoms couple to the vacuum modes, they introduce additional noise into the system.
The corresponding term reads
\begin{equation}
  \label{addNoise}
  \X_{\overline{\Phi}} = \frac{w}{\gnu+w}\left[\X_\Pb + \frac{\gnu}{g}\left(\X_\Sb - \frac{\Delta}{\Gamma}\X_\Lb\right)\right]\,.
\end{equation}

As already mentioned, we have to calculate the value of the single-atom population inversion $z_M$ in a consistent way.
Therefore, we take the expectation values of Eqs.~\eqref{HLEactive}, solve them in the steady state, and impose the condition
\begin{equation}
  \overline{P} = \frac{1+z}{2}\,.
\end{equation}
This yields
\begin{align}
  \label{zM}
  z_M = & \frac{1}{2M w(\gnu)}\biglb(\kappa(\gnu+w) + M w(\nu-\gamma) \nonumber\\
    & - \bigl\{4\kappa w(\gnu)M(\gamma-\nu+w)\nonumber\\
    & + \left[\kappa(\gnu+w) + M w(\nu-\gamma)\right]^2\bigr\}^{1/2}\bigrb)\,.
\end{align}
Note that from Eq.~\eqref{zM} it is straight forward to show that $\zeta < 1$.

\subsection{Atomic motion in a hot cavity}
\label{subsec:motion}
Let us now investigate the light forces inside the active resonator by adding a single additional active atom at position $x$ with the position dependent coupling $G = g \cos(k x)$.
In some previous work, we have investigated such a configuration without the other atoms present, where the atom itself has to provide for the entire gain.
Even then, lasing, cooling, and trapping can coexist for a blue detuned light field \cite{SALletter} under suitable conditions.
We found that atomic equilibrium temperatures below the Doppler limit are possible, but this poses very stringent conditions onto the atomic pumping mechanisms.
Characteristically, the final atomic equilibrium temperature decreases with higher intracavity photon numbers and the particle's localization is enhanced.

Here we consider already $M$ active atoms present in the resonator which contribute the major part of the cavity photon number.
Hence, we can expect to get a deeper optical potential and faster mechanical cooling rates.
Of course, as inevitable for any quantum consistent amplification mechanism, the atoms representing the active medium will introduce additional noise in the field.
This results in enlarged fluctuations of the system-operator expectation values, which in turn will imply also stronger momentum diffusion.
The combined effect of these two contributions will be worked out in the following.

As the total population inversion needed for crossing the laser threshold is now shared among many atoms, we get less stringent pumping requirements and a strongly reduced dependence of the cavity field on the individual atomic positions as compared to the single-atom (passive-cavity) case.
In particular, there will be still photons and thus an optical potential present, even when the extra atom is located very close to a field node.

Generally, we expect that larger photon numbers lead to lower temperatures since cooling relies on energy dissipation by photons that leave the resonator via its mirrors.
In the ideal regime, each photon carries away some atomic kinetic energy.
As a second advantage of more atoms, we expect that heating due to spontaneous emission is diminished as the atoms share the required excitation.
Finally, since the equilibrium temperature for cavity cooling scales with the linewidth $\kappa$ \cite{Horak}, an effectively gain-reduced linewidth lets us further expect enhanced cooling.

In what follows we consider a single atom moving under the influence of the field created by stimulated emission of the atom itself and the active medium described in Sec.~\ref{subsec:active}.
We therefore have to add the contribution of the moving atom to the Hamiltonian \eqref{H}.
Writing $\Pi=\sigma_+\sigma_-$, $\Sigma=a^\dagger\sigma_-+\sigma_+ a$, and $\Lambda=-i(a^\dagger\sigma_--\sigma_+ a)$, with $G(x) = g f(x)$, we have 
\begin{equation}
  \H = \H_\mathrm{int} - \Delta\Pi - G(x)\Lambda\,.
\end{equation}
Using again the factorization approximation $(2\Pi-1)\Phi \approx Z\Phi$, the corresponding linearized HLE for the field and the extra atom read
\begin{subequations}
  \label{HLEsingle}
  \begin{align}
    \dot{\Phi} &= -2\keff\Phi + G(x)\Sigma + 2M\eta + \X_\Phi + \X_{\overline{\Phi}}\,,\\
    \dot{\Pi} &= -2\gamma\Pi - G(x)\Sigma + 2\nu(1-\Pi) + \X_\Pi\,,\\
    \dot{\Sigma} &= -\Gamma\Sigma - \Delta\Lambda + 2G(x)Z\Phi + 2G(x)\Pi + \X_\Sigma\,,\label{singleSigma}\\
    \dot{\Lambda} &= -\Gamma\Lambda + \Delta\Sigma + \X_\Lambda\,.
  \end{align}
\end{subequations}
These equations, containing quantum correlations up to second order in the field amplitude and atomic polarizations, will be the basis of all our further considerations.

\subsection{Photon number and forces in the steady state} 
\label{subsec:steady}
As already mentioned above, as a crucial step to obtain an explicit approximate solution, we replace the population inversion operators $\sigma_{z,m}$ and $\sigma_z = \sigma_{+}\sigma_{-}-\sigma_{-}\sigma_{+}$ by their expectation values in the nonlinear operator products of Eqs.~\eqref{activeSigma} and \eqref{singleSigma} involving the photon number operator $\Phi$.
In practice we proceed as follows.
We first determine $z_{M+1}$ from Eq.~\eqref{zM} by assuming that all $M+1$ atoms are fixed at antinodes.
This population inversion $z_{M+1}$ then fixes the effective linewidth $\kappa_\mathrm{eff}$ and we can get the population inversion of the moving atom by solving for $\langle\Pi\rangle = (Z+1)/2$ in the steady state.
After some algebra we get
\begin{align}
  Z = & \frac{1}{2W(\gnu)}\biglb(\keff(\gnu+W) + W(\nu-\gamma + 2M\eta) \nonumber\\
  & - \bigl\{4\keff W(\gnu)(\gamma-\nu+W) \nonumber\\
  & + \left[\keff(\gnu+W) + W(\nu-\gamma+2M\eta)\right]^2\bigr\}^{1/2}\bigrb)\,.
\end{align}
Note that this approximation holds well as long as the extra atom is near a field antinode, where the spatial variation of the coupling strength is small.
If the motion of the atom is sufficiently slow, the dynamics of the internal variables can adiabatically follow any atomic displacement and we can approximate the state of the system by the steady state for an atom at fixed position.
This allows to calculate the average force acting on the atom (dipole force) from the expectation value of the force operator, given by the spatial derivative of the interaction term in the Hamiltonian \cite{Cohen},
\begin{equation}
  \label{forceOperator}
  \hat{F} = -\nabla\mathcal{H} = (\nabla G)\Lambda\,.
\end{equation}

Since any photon in the cavity originates from stimulated emission, the stationary photon number $N$ should directly be related to the atomic excitation.
Indeed, by including the contribution of the moving atom, we find a relation analogous to Eq.~\eqref{relColl},
\begin{equation}
  \label{relNew}
  M\nu(1-\overline{P})  + \nu(1-P) = \kappa N + M\gamma\overline{P} + \gamma P\,.
\end{equation}
Notice that this equation reduces to Eq.~\eqref{relColl}, when the coupling of the atom to the lasing mode vanishes, i.e.\ at field nodes.
Figure \ref{fig:drei} (a) shows the stationary photon number per atom $n = N/(M+1)$ as a function of the pumping rate $\nu$ for $(\gamma,\Delta,g) = (10,40,4)\kappa$ when the atom is fixed at an antinode.
We plot here only the range $\nu>\gamma$ since lower pumping strengths would prevent population inversion.
Each curve exhibits a threshold value for $\nu$  which evidently approaches $\gamma$ with growing $M$.
As one would further expect for our lasing system, the light intensity increases linearly with $\nu$ above threshold.
Yet, the fact that $n$ does not remain constant for increasing $M$ but increases as well, demonstrates the cooperative emission properties of the system.
It should therefore be possible to achieve higher gain parameters by incorporating many atoms for stimulated emission.
This behavior is illustrated in Fig.~\ref{fig:drei} (b).
It depicts the corresponding effective resonator linewidth $\keff$ that is directly related to the gain parameter $\zeta$ via Eq.~\eqref{kappaEff}.
For $\nu > \gamma$, the effective resonator linewidth is reduced linearly with $\nu$ until the system passes threshold.
There, the population inversion and hence $\keff$ have already reached their minimum values and only slightly decrease.
As this happens close to threshold, involving many atoms implies enhanced gain for fixed parameters.
We can therefore hope for improved cavity cooling since the final atomic temperature scales with the resonator linewidth there.

\begin{figure}
  \includegraphics[width=8cm]{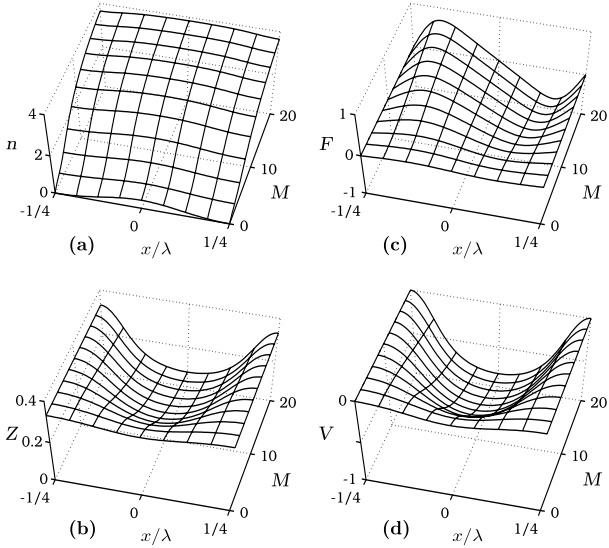}
  \caption{Dependence of (a) the photon number per atom $n$, (b) the atomic population $Z$, (c) the force $F$, and (d) the light potential $V$ on the position $x$ of a single moving atom for varying atom number $M$ that constitute the gain medium.
    The parameters are the same as in Fig.~\ref{fig:zwei}.}
  \label{fig:vier}
\end{figure}

In Fig.~\ref{fig:vier} (a) and (b) we show the dependence of, respectively, $n$ and $Z$ on the atomic position within half a wavelength for different atom numbers $M$.
The change in the corresponding light potential [c.f.\ Fig.~\ref{fig:vier} (d)] results in a net force acting on the atom which can be calculated from the steady-state solution of Eqs.~\eqref{HLEsingle}.
It is given by
\begin{equation}
  \label{df}
    F = \frac{\keff\nu+M\eta(\gnu)Z}{\Deff}\frac{2\Delta W}{G}(\nabla G)\,,
\end{equation}
where we have defined the determinant of the linear system,
\begin{equation}
  \Deff = \keff(\gnu+W) -(\gnu)W Z\,.
\end{equation}
From Eq.~\eqref{df} we see that the dipole force is an odd function of the atom-field detuning $\Delta$.
In analogy to the single-atom case \cite{SALletter}, the atom is pushed towards regions with large intensity (high field seeker) for a blue detuned laser field.
A plot of $F$ for $\Delta > 0$ can be found in Fig.~\ref{fig:vier} (c).

When the atom is confined to the vicinity of field antinodes, the spatial variations in the quantities $W$ and $Z$ drop.
In this case, the force mediated by the active medium with $M \approx M+1 \gg 1$ atoms, in units of the single-atom force $F_0$, is given by
\begin{equation}
  \label{Fenhance}
  \frac{F}{F_0} = \frac{\kappa}{\keff} = \frac{1}{1-\zeta}\,.
\end{equation}
This result indicates a gain-induced enhancement of the local forces close to field antinodes which, in general, are small there due to low field gradients.
The increase of the dipole force and hence the light potential leads to strong localization of the atom.
It should be emphasized that localization effects become particularly important for the velocity-dependent (friction) forces within the scope of the cooling performance.
However, let us refer at this point to the discussion at the end of Sec.~\ref{subsec:temperature}.

\subsection{Atomic temperature and localization}
\label{subsec:temperature}
Let us now drop the assumption of slow atoms and investigate atomic motion in more detail.
In particular, we calculate the particle's equilibrium temperature $T$ associated with its COM motion.
Comparing $k_\mathrm{B}T$ to the optical potential depth, we then get information about localization effects.
Using the well known Einstein relation \cite{Cohen}, the temperature of the atom can be estimated by the ratio of the spatially averaged momentum diffusion term $\mathcal{D}$ and the linear friction coefficient $\beta$,
\begin{equation}
  k_\mathrm{B} T = \overline{\mathcal{D}}/\overline{\beta}\,.
\end{equation}
Following the procedure of Ref.~\cite{SALarticle}, we can obtain both of these quantities from the force operator \eqref{forceOperator}.

We will not present the whole calculation here apart from mentioning the basic ideas.
As long as the atom moves much less than a wavelength within the characteristic time scale of the internal dynamics, the friction can be approximated by the force term linear in the particle's velocity $v$.
A simultaneous expansion of both Eqs.~\eqref{HLEsingle} and the system operators in terms of $v$ results in a set of dynamical equations, where the first order solution yields the friction coefficient
\begin{equation}
  \beta = (\nabla G)\langle\Lambda^1\rangle\,.
\end{equation}
The full expression of $\langle\Lambda^1\rangle$ is listed in App.~\ref{app:fric}.

For the diffusion coefficient we have to evaluate the two-time correlation function of the force operator \cite{Gordon} since momentum diffusion arises from fluctuations of this operator.
Here we will use the approach of Ref.~\cite{Domokos} as it allows us to directly relate momentum diffusion to the noise caused by spontaneous emission and cavity decay.
We therefore read off $\mathcal{D}$ from
\begin{equation}
  \label{eq:diffDef}
  \langle\hat{F}(t)\hat{F}(t-\tau)\rangle - \langle\hat{F}(t)\rangle\langle\hat{F}(t-\tau)\rangle = 2\mathcal{D}\delta(\tau)\,.
\end{equation}
A very brief derivation can be found in App.~\ref{app:diff}.
We also have to include heating from the random recoil of spontaneously emitted photons, which results in an additional diffusion term that can be found in Ref.~\cite{Cohen}.

\begin{figure}
  \includegraphics[width=8cm]{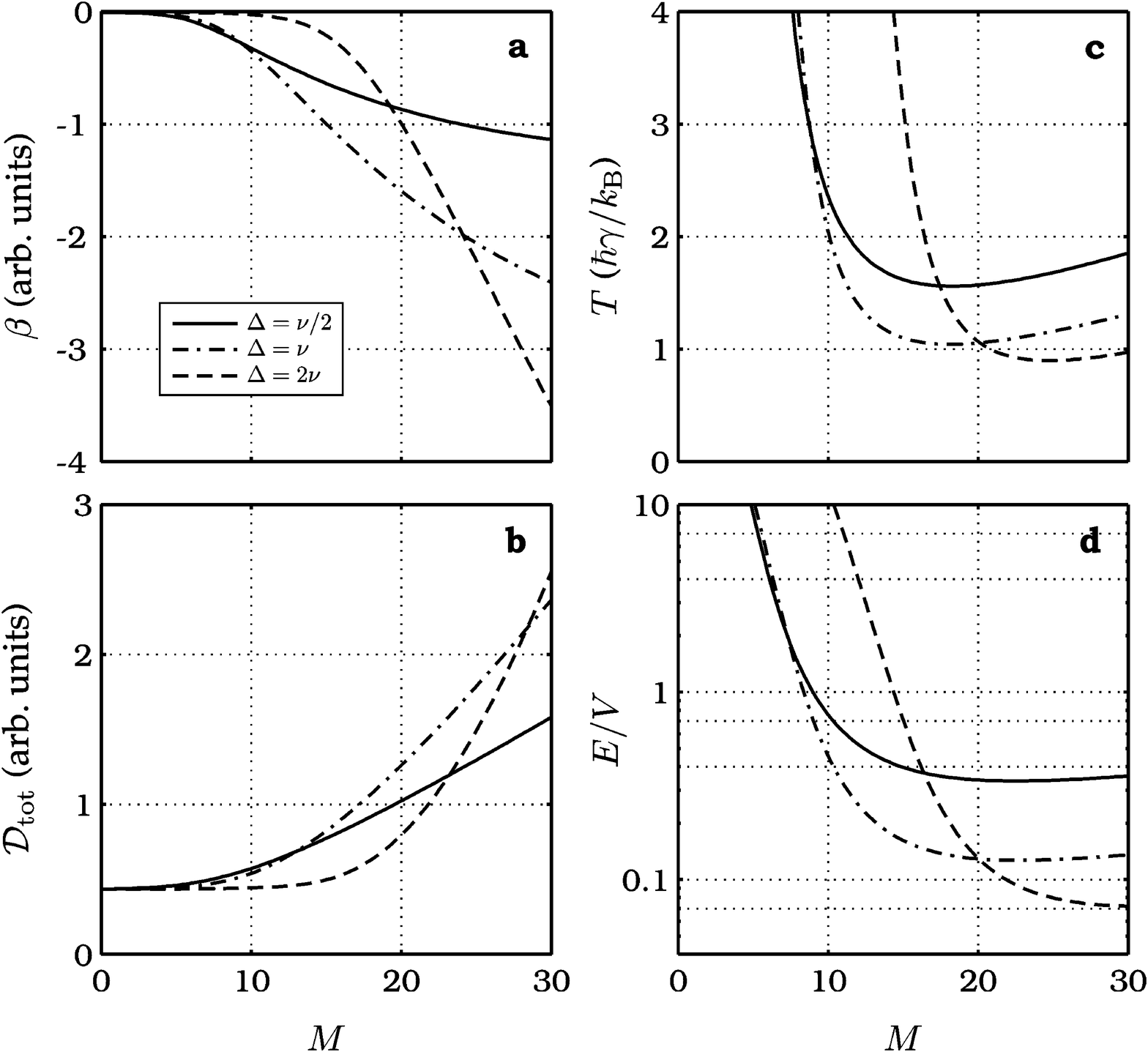}
  \caption{(a) Position averaged friction coefficient $\beta$ in arbitrary units vs $M$ for $(\gamma,\nu,g)=(10,20,4)\kappa$ and different values of $\Delta$, as indicated.\\
  (b) Average of the total diffusion coefficient $\mathcal{D}_\mathrm{tot}$ for the same parameters.\\
  (c) Approximate atomic equilibrium temperature in units of the Doppler temperature.\\
  (d) Ratio of the kinetic energy and the light potential as an indicator for localization of the particle.}
  \label{fig:funf}
\end{figure}

In Fig.~\ref{fig:funf} (a) and (b) we depict the scaling of, respectively, the position averaged friction and total diffusion coefficients with the amount of atoms in the active medium, $M$, for different values of the detuning $\Delta$.
(Note that we plot continuous curves instead of discrete values here, and that $M = 0$ corresponds to the one-atom case.)
We chose $\nu = 2\gamma$ in order to ensure population inversion such that $\beta$ is negative.
Clearly, we see that both $\beta$ and $\mathcal{D}_\mathrm{tot}$ strongly depend on the number of atoms, and their absolute values grow with $M$.
Since each additional atom increases the dipole force $F$ and thus the light potential $V$ [see Fig.~\ref{fig:vier} (c) and (d)], one could expect this effect to arise from enlarged light intensities when many atoms constitute the medium.
Indeed, due to the fact that friction relies on the time-delayed force caused by an atomic motional displacement, we should conclude that $\beta$ shows similar scaling properties with $M$ as the dipole force $F$.
On the other hand, the fluctuating part of the force operator quantifies momentum diffusion, not its steady-state value.
As we saw from Eq.~\eqref{singleDiff}, the active medium introduces an extra noise term which should lead to additional momentum diffusion.
Moreover, we infer from Fig.~\ref{fig:funf} (a) and (b) that larger values of the atom-field detuning $\Delta$, certainly decreasing the photon number, result in larger friction as well as momentum diffusion.
However, we will later see how we can compare situations with different atom numbers in a more favorable way where the photon number, the atomic population inversion, and the gain parameter are independent of $M$.
Let us finally mention here that the cooling rate is inversely proportional to $\beta$ such that large values of $\beta$ will lead to fast cooling.

The resulting equilibrium temperature $T$ associated with the atomic COM motion is shown in Fig.~\ref{fig:funf} (c) in units of the Doppler temperature $T_\mathrm{D} = \hbar\gamma/k_\mathrm{B}$.
Analyzing this plot, we can identify two regimes according to the scaled pumping rate $y = \nu/\Delta$.
As long as there are only a few atoms in the cavity, the atomic temperature decreases with increasing $y$.
In fact, this is in strong analogy to the single-atom case where we showed that temperatures even below the limit of passive cavity cooling are possible for sufficiently large $y$ \cite{SALarticle}.
Here, on the other hand, the addition of even a single atom can lead to a significant drop in the temperature, which already proves the importance of cooperative aspects.
Depending on the actual value of the detuning $\Delta$, we eventually find, with growing $M$, a transition point to a second regime where the cooling performance can be enhanced for lower pumping rates $y$ that we achieved by raising $\Delta$.
The cooling enhancement only happens for sufficiently large atom numbers and is directly related with the fact that strong pumping immediately saturates the gain here.
Indeed, one has to ensure not to saturate the gain as the cooperative effects of emission will then break and, as a consequence, the cooling efficiency is reduced.
Each curve shows a minimum which tends towards larger $M$ when we increase the detuning $\Delta$.
For fixed parameters, the minimum corresponds to a somehow optimum atom number $M$ which keeps the balance between the required population inversion on the one hand and saturation on the other hand.
In this sense, lowering the scaled pumping rate $y$ means that we can involve more atoms into the cooling process without saturating the gain at once.
Note that for the chosen parameters here, which do not require to be in the strong atom-field coupling regime, one can achieve sub-Doppler cooling for $M \gtrsim 20$. 

In view of a setup where one has to assure that the atoms remain in the interaction region close to field antinodes, we will now focus on localization effects.
Therefore we plot in Fig.~\ref{fig:funf} (d) the ratio of the atomic kinetic energy $E$ and the light potential $V$.
Large values of this ratio correspond to a more or less even distribution of the particle's position while values below one indicate strong localization.
One can again identify the afore mentioned regimes of low and high atom numbers $M$, respectively.
Qualitatively, the curves for $E/V$ are similar to the ones for the temperature and we recognize that cooling is directly accompanied by atomic trapping for sufficiently high $M$.
In particular, the ratio $E/V$ drops well below unity when the atomic temperature gets down into the range of the Doppler limit.

As already mentioned before, the cooling properties may significantly depend on the initial atomic temperature and localization.
For rather cold and well localized atoms, in particular, the position dependence of the cooling forces must be taken into account.
Typically, the cooling time scale strongly increases as soon as an initially rather hot ensemble of cavity-cooled atoms gets trapped.
(Notice, that we have already seen a similar effect in Sec.~\ref{sec:semi} although no information about the equilibrium temperature and cooling time could be given there.)
Moreover, it has been shown that in the limit of very strong confinement, the friction forces can even vanish such that the cooling process completely stops \cite{gangl:rapid}.
On the other hand, strong cooling forces may arise in the high-intensity regime resulting in very strong atomic localization \cite{semiclassical,selforg}, where the additional momentum diffusion leads to a mere slight increase of the atomic equilibrium temperature.

In order to get an estimate of the influence of localization effects on the cooling performance in a hot cavity, we will derive the gain-induced enhancement of the friction for a rather well localized atom.
For $M \approx M+1 \gg 1$, it is given by
\begin{equation}
  \label{betaenhance}
  \frac{\beta}{\beta_0} = \left(\frac{\kappa}{\keff}\right)^3 = \left(\frac{1}{1-\zeta}\right)^3\,,
\end{equation}
where $\beta_0$ indicates the single-atom friction coefficient.
In the previous section, we have calculated the same expression for die stationary force which turned out to scale inversely with the effective linewidth $\keff$.
Since $\beta/\beta_0 \sim \keff^{-3}$, the cooling forces are even more enhanced than the trapping potential such that an atom trapped very close to field antinodes still experience relatively large friction as soon as $1 - \zeta \ll 1$.
We therefore deduce that, besides providing strong particle localization, the stimulated-emission mediated cooling forces can significantly accelerate the cooling process as compared to conventional schemes using passive resonators.

\begin{figure}
  \includegraphics[width=8cm]{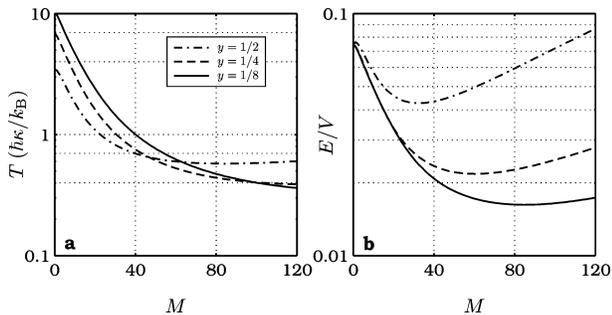}
  \caption{(a) Equilibrium temperature $T$ in units of the minimum temperature achievable by passive cavity cooling.
    $T$ is plotted as a function of the atom number $M$ where $\kappa/(M+1)$ was kept constant for different pumping rates $y$, as indicated.\\
  (b) Corresponding ratio $E/V$.}
  \label{fig:sechs}
\end{figure}

\subsection{Scaling properties with the number of atoms}
\label{subsec:scaling}
In the good-cavity limit ($\kappa \ll \nu$), the gain-induced cooling of a single particle can lead to atomic temperatures well below the limit of passive cavity cooling for sufficiently large amount of pumping ratio $y$ and when one can neglect spontaneous emission \cite{SALarticle}.
Nonetheless, the required large excitation rate $\nu$ certainly poses a serious experimental challenge there.
As we already have seen from Fig.~\ref{fig:funf}, the use of many atoms for stimulated photon emission into the resonator can enhance trapping and also lower the particle's kinetic energy, even for small values of $y$.
In the following, we will now analyze the scaling of the cooling properties with respect to the atom number $M$ to see whether this method should work for larger ensembles under less restrictive experimental conditions.

For this purpose, we introduce a rescaling of the system parameters as follows.
As we have seen above, the comparison of the motional properties for different $M$ is actually not trivial.
Particularly, not only the photon number $N$ but also the population inversion $z_M$ as well as the gain parameter $\zeta$ play a crucial role there as they all depend on $M$ in a highly nonlinear way.
A good choice will certainly be a rescaling which keeps these three quantities constant at all.
A closer look at Eqs.~\eqref{zM}, \eqref{zeta}, and \eqref{singleDiff} reveals that one can realize this all at once by keeping the quotient $\kappa/(M+1)$ constant.
Consequently, the operating point of the lasing system can be set independently from the atom number without any further complications.
We will accomplish this here by relating the total amount of emission into the resonator to the photon loss rate according to
\begin{equation}
  (M+1) w = a \kappa
\end{equation}
via the parameter $a$.
The value of $a$ and hence the operating point can then be set by a proper choice of the coupling constant $g$.

Besides, this rescaling accounts for comparable conditions also in the energy balance.
Having fixed a certain operating point, the established light intensity no longer depends on the number of atoms involved.
Regardless, however, the rate of photon leakage through the cavity mirrors, given by $2\kappa N$, linearly grows now with the atom number.
Since each photon carries away energy and entropy from the system, we expect not only lower temperatures but also faster cooling.

In Fig.~\ref{fig:sechs} (a) we show the scaling of the atomic temperature with $M$ for different values of $y$ where we assumed $\gamma = 0$.
Clearly, $T$ drops well below the limit of passive cavity cooling which is given by $k_\mathrm{B}T_\mathrm{cav} = \hbar\kappa$.
When many atoms are involved, this happens even for values of the pumping ratio $y$ beneath the single-atom limit $y = 1/2$.
The corresponding ratio $E/V$ is depicted in Fig.~\ref{fig:sechs} (b).
Again we find simultaneous trapping and cooling.
This result suggests the possibility to use a rather bad cavity and many weakly pumped active atoms as gain medium to achieve stimulated emission induced cooling.
Fortunately, these conditions simultaneously ease the requirements on cavity technology and on the necessary amount of pumping per atom $y$ and give a rather promising perspective of possible implementations of the scheme.

\section{conclusions}
We have shown that intracavity gain significantly modifies light forces in cavities.
In particular, the stimulated-emission dominated interaction of individual particles with a single light mode opens new perpectives on the dynamics of lasing and light-induced manipulation of the atomic motion.
To exhibit the underlying physics in detail, we developed a quantum consistent theory for intracavity gain, based on a microscopic description of an ensemble of individual inverted atoms strongly coupled to a single resonator mode.
Including light forces into well-established laser equations, we found a self-consistent analytical model for the combined dynamics of the atomic motion in a hot cavity.
As an extra bonus, by accounting for spontaneous emission and cavity decay, fluctuations and saturation are automatically included in our model by construction.

As a central result, we can confirm previous claims \cite{Vuletic} that laser cooling can be strongly improved in such a setup with less stringend requirements on cavity technology.
This result turns out to be closely related to a strong reduction of the effective resonator linewidth as the amplifying atoms partly balance the photon loss via stimulated emission.
In this way, photon scattering into the cavity, which accounts for the cooling forces, can be drastically enhanced. 
While the naive expectation of a much reduced temperature due to the narrow linewidth is not fully met by the results, we still get much faster cooling and sub-Doppler temperatures.
Our findings show the potential usefulness of stimulated-emission induced cooling in optical resonators and may open the way to novel experiments.

More importantly however, we recover that optical lasing can be concurrent with strong atomic localization and sub-Doppler cooling.
This directly points to the possibility of combined photon lasing and atom lasing in the continuous-wave (CW) regime.
In a conceivable joint setup, a coherent atomic beam could be out-coupled from an ultra cold ensemble of atoms in an optical lattice.
The atoms are cooled to the motional ground state without particle loss by Raman gain involving stimulated emission into a resonator mode.
The photon leakage through the cavity mirrors provides the photon laser channel there.
In order to achieve combined CW lasing operation, the cooling region is fed by atomic tunneling from neighbouring sites.
Ground-state atoms which are prepared in a non coupling Hyperfine state by a proper microwave beam leave the trap due to gravity.


\acknowledgments
The authors would like to thank P.\ Domokos for fruitful discussions.
This work was supported by the Austrian Science Foundation FWF under projects P17709N08 and the SFB ``Control and Measurement of Coherent Quantum Systems''.

\appendix

\section{Friction force}
\label{app:fric}
We have
  \begin{align}
    \langle\Lambda^1\rangle = & -\frac{\Delta W^2(\nabla Z)}{\Gamma^2 G^3\Deff^3}\left[\nu W + M\eta(\gnu+W)\right]\nonumber\\
      & \times\bigl\{4\keff^2(\gnu)^2\Gamma + G^2\bigl[\keff^2(\gnu-\kappa)\nonumber\\
      & + (\gnu)^2(\Gamma-2\keff)Z\bigr]\bigr\}\nonumber\\
      & -\frac{\Delta W(\nabla W)}{\Gamma^2 G^3\Deff^3}\left[\keff\nu + M\eta(\gnu)Z\right]\nonumber\\
      & \times\biglb(2\keff(\gnu)\Gamma\bigl[\keff(\gnu-W)\nonumber\\
      & + (\gnu)W Z\bigr] - G^2\bigl\{\kappa\keff^2-(\gnu)^2\nonumber\\
      & \times(\Gamma-\keff)Z + W\left[\keff-\left(\gnu\right)Z\right]^2\bigr\}\bigrb)\,.
  \end{align}

\section{Diffusion coefficient}
\label{app:diff}
Writing $\hat{F} = \langle\hat{F}\rangle + \X$, Eq.~\eqref{eq:diffDef} becomes
\begin{equation}
  \label{app:corr}
  \langle\X(t)\X(t-\tau)\rangle = 2{\cal D}\delta(\tau)\,,
\end{equation}
where the fluctuating part of the force operator reads for quasi-stationary conditions and for $D=(\Gamma^2+\Delta^2)\Deff$
  \begin{align}
    \label{X}
    \X = & \frac{(\nabla G)}{D}\bigl\{(\gnu)\Delta G Z\left(\X_\Phi + \X_{\overline{\Phi}}\right)\nonumber\\
      & + \keff\Delta G\X_\Pi + \keff(\gnu)\left(\Delta\X_\Sigma + \Gamma\X_\Lambda\right)\nonumber\\
      & + G^2[\keff-(\gnu)Z]\X_\Lambda\bigr\}\,.
  \end{align}
Evaluating the correlation function \eqref{app:corr}, one finds for the contribution of the active medium to the diffusion coefficient
\begin{align}
  \label{app:diffActive}
  \mathcal{D}_\mathrm{act} = & \frac{(\gnu)^2\Delta^2 G^2Z^2(\nabla G)^2}{D^2}\frac{M w}{\Gamma(\gnu+w)^2}\nonumber\\
    & \times\bigl\{(\gnu)^3\langle\Phi\rangle + \left[\kappa(\gnu)^2+(\gamma-\nu)w\Gamma\right]\langle\Pb\rangle\nonumber\\
    & + (\gnu)\Gamma/g \left[\kappa(\gnu+w)+w(\gamma-\nu)\right]\nonumber\\
    & \times\left(\Gamma^2-\Delta^2\right)/\Gamma^2\langle\Sb\rangle + \nu\left[w\Gamma+(\gnu)^2\right]\bigr\}
\end{align}
by collecting all terms involving the operator $\X_{\overline{\Phi}}$.
Here we used the steady-state solution of the collective operators
\begin{subequations}
  \begin{gather}
    \langle\Pb\rangle = \frac{\nu-w z_{M+1}\langle\Phi\rangle}{\gnu+w}\,,\\
    \langle\Sb\rangle = \frac{2w}{g}\frac{\nu+(\gnu)z_{M+1}\langle\Phi\rangle}{\gnu+w}\,,\\
    \langle\Lb\rangle = \frac{\Delta}{\Gamma}\langle\Sb\rangle\,.
  \end{gather}
\end{subequations}
The remaining portion finally reads
\begin{align}
  \mathcal{D}_\mathrm{sin} = & \frac{(\nabla G)^2}{\Deff D^2}\biglb(\kappa (\gnu)^2\Delta^2 G^2 Z^2\bigl[\nu W \nonumber\\
    & + M\eta(\gnu+W)\bigr] + \keff^2\Delta^2 G^2\nonumber\\
    & \times\left[\keff W+2\gamma(\keff-W Z) + M\eta(\nu-\gamma)W Z\right]\nonumber\\
    & + 2W\Delta^2\left[\keff\nu + M\eta(\gnu)Z\right]\nonumber\\
    & \times\left[\keff(\gamma-\nu) + \kappa(\gnu)Z\right] \nonumber\\
    & \times\left\{2\keff(\gnu) + \left[\keff-(\gnu)Z\right]G^2/\Gamma\right\}\nonumber\\
    & + \bigl\{\nu\left(\keff\Gamma + \Gamma_\mathrm{eff}W - \Gamma W Z\right) \nonumber\\
    & + M\eta\left[(\gnu)(\gnu+W)-\kappa W Z\right]\bigr\} \nonumber\\
    & \times\bigl\{(\Gamma^2+\Delta^2)\keff^2(\gnu)^2 + G^4\left[\keff-(\gnu)Z\right]^2 \nonumber\\
    & + 2\keff(\gnu)\Gamma G^2[\keff-(\gnu)Z]\bigr\}\bigrb)\,.
\end{align}

\end{document}